\documentclass[journal]{IEEEtran}
\usepackage{amsmath,amsfonts,bm}

\def\eqref#1{equation~\ref{#1}}
\def\1{\bm{1}}

\def\vc{{\bm{c}}}

\def\ve{{\bm{e}}}

\def\vu{{\bm{u}}}
\def\vv{{\bm{v}}}

\def\vx{{\bm{x}}}
\def\vy{{\bm{y}}}

\def\mA{{\bm{A}}}
\def\mB{{\bm{B}}}

\def\mE{{\bm{E}}}

\def\mH{{\bm{H}}}
\def\mI{{\bm{I}}}

\def\mW{{\bm{W}}}
\def\mX{{\bm{X}}}

\DeclareMathAlphabet{\mathsfit}{\encodingdefault}{\sfdefault}{m}{sl}
\SetMathAlphabet{\mathsfit}{bold}{\encodingdefault}{\sfdefault}{bx}{n}

\def\sR{{\mathbb{R}}}

\usepackage{booktabs,multirow}
\usepackage{color}
\usepackage{makecell}
\usepackage{amsmath}
\usepackage{hyperref}
\usepackage[table]{xcolor}
\ifCLASSINFOpdf
\else
   \usepackage[dvips]{graphicx}
\fi
\usepackage{url}

\usepackage{graphicx}

\begin{document}

\title{Enhancing Intelligibility for Generative Target Speech Extraction via Joint Optimization with Target Speaker ASR}

\author{Hao Ma, Rujin Chen, Xiao-Lei Zhang, Ju Liu, and Xuelong Li, \textit{Fellow, IEEE}
\thanks{Hao Ma and Ju Liu are with the School of Information Science and Engineering, Shandong University, Qingdao, China; and Hao Ma is also with the Institute of Artificial Intelligence (TeleAI), China Telecom. (e-mail: mahao2000818@hotmail.com, juliu@sdu.edu.cn).}
\thanks{Rujin Chen, Xiao-Lei Zhang, and Xuelong Li are with the Institute of Artificial Intelligence (TeleAI), China Telecom. (e-mails: preciouscrj@163.com, xiaolei.zhang@nwpu.edu.cn, xuelong\_li@ieee.org).}
% \thanks{Ju Liu is with the School of Information Science and Engineering, Shandong University, Qingdao, China. (e-mail: juliu@sdu.edu.cn)}
% \thanks{Corresponding Author: Ju Liu.}
\thanks{Work was done during Hao Ma's internship at TeleAI. Corresponding authors: Ju Liu and Xiao-Lei Zhang.}
}

\markboth{Journal of \LaTeX\ Class Files, Vol. 14, No. 8, August 2015}
{Shell \MakeLowercase{\textit{et al.}}: Bare Demo of IEEEtran.cls for IEEE Journals}
\maketitle

\begin{abstract}
Target speech extraction (TSE) isolates the speech of a specific speaker from a multi-talker overlapped speech mixture.
Most existing TSE models rely on discriminative methods, typically predicting a time-frequency spectrogram mask for the target speech. However, imperfections in these masks often result in over-/under-suppression of target/non-target speech, degrading perceptual quality.
Generative methods, by contrast, re-synthesize target speech based on the mixture and target speaker cues, achieving superior perceptual quality. Nevertheless, these methods often overlook speech intelligibility, leading to alterations or loss of semantic content in the re-synthesized speech.
Inspired by the Whisper model's success in target speaker ASR, we propose a generative TSE framework based on the pre-trained Whisper model to address the above issues. This framework integrates semantic modeling with flow-based acoustic modeling to achieve both high intelligibility and perceptual quality.
Results from multiple benchmarks demonstrate that the proposed method outperforms existing generative and discriminative baselines. We present speech samples on https://aisaka0v0.github.io/GenerativeTSE\_demo/.
\end{abstract}

\begin{IEEEkeywords}
Generative target speech extraction, target speaker extraction, Whisper, optimal transport conditional flow matching, multi-task joint learning
\end{IEEEkeywords}

\IEEEpeerreviewmaketitle

\section{Introduction}

\IEEEPARstart{H}{umans} possess a remarkable ability to focus on specific speech in noisy environments, a phenomenon known as the \textit{cocktail party effect}. In signal processing, speech separation \cite{SS} has been extensively studied to address this challenge by decomposing speech mixtures into independent sources. Target speech extraction (TSE), in contrast, focuses on isolating the speech of a specific target speaker by leveraging enrollment cues associated with that speaker. This approach holds significant potential for various real-world applications.

{Existing research on TSE can be categorized into \textit{discriminative} \cite{speakerbeam_lstm, speakerbeam_cnn, xu2020spex} and \textit{generative} methods \cite{difftse, generative, tang2024tselm}.}
Discriminative methods directly minimize the distance between the model's estimation and the target speech, typically by predicting a target speech mask and applying it to the time-frequency spectrogram of the speech mixture. 
{Following this paradigm, significant efforts have been devoted to using various neural network architectures for accurate target speech mask modeling.}
{Early studies employed convolution neural networks \cite{speakerbeam_cnn} or recurrent neural networks \cite{speakerbeam_lstm} to model target speech masks. More recently, advanced architectures such as Transformers \cite{xsep} and novel techniques like the band-split method \cite{pbsrnn} have further improved the mask modeling accuracy. %Target speaker cues are typically introduced as a global speaker embedding from the enrollment speech. 
Some recent studies \cite{zhang2024multi, multi_1} have explored incorporating finer-grained target speaker cues to reduce the risk of mistakenly extracting non-target speech.}
% Early works employed convolution neural networks \cite{speakerbeam_cnn} or recurrent neural networks \cite{speakerbeam_lstm} for target speech mask modeling. More recent works have improved modeling accuracy through advanced architectures like Transformers \cite{xsep} or novel techniques such as the band-split module \cite{pbsrnn}. Additionally, some studies \cite{zhang2024multi, multi_1} have explored incorporating finer-grained target speaker cues to reduce the risk of mistakenly extracting non-target speech.
Despite these advancements, mask-based discriminative methods inherently suffer from imperfections in mask modeling. Such limitations can lead to over-/under-suppression of target/non-target speech, ultimately degrading perceptual quality.%—an intrinsic challenge for this approach.

\begin{figure}
    \centering
    \includegraphics[width=0.8\linewidth]{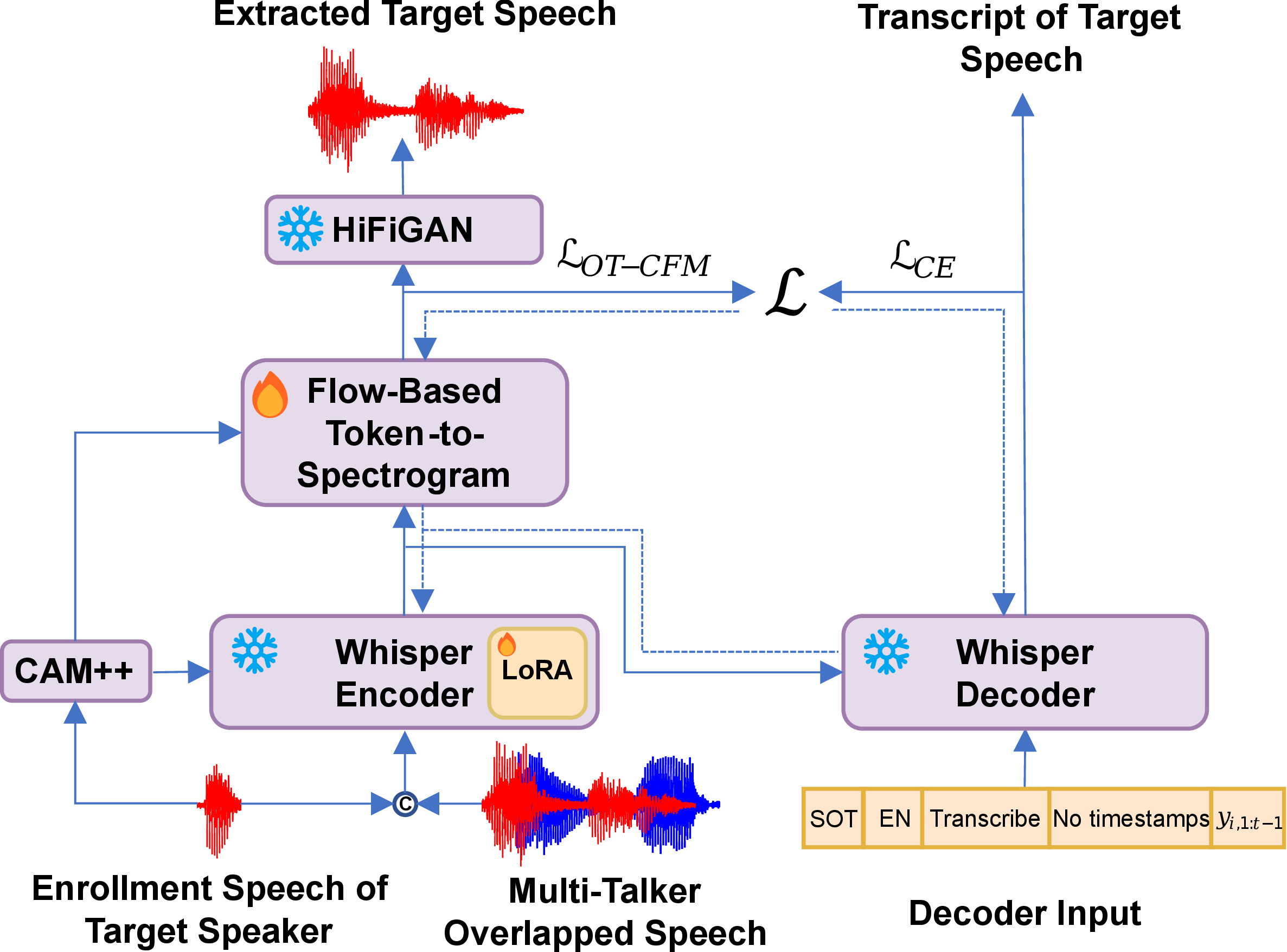}
    \vspace{-2mm}
    \caption{Overview of the proposed method. The solid line represents the forward propagation flow; the dashed line represents the backpropagation flow.}
    \vspace{-6mm}
    \label{fig:main}
\end{figure}

On the other hand, generative methods directly model the distribution of the clean speech for TSE, resulting in improved perceptual quality \cite{generative}.
% Recent advancements in speech enhancement \cite{selm, masksr}, separation \cite{diffse}, and TSE \cite{difftse, generative, tang2024tselm} have made progress from various perspectives.
Specifically, \cite{difftse} introduces a conditional diffusion model for generative TSE. In \cite{generative, tang2024tselm}, researchers focus on the discrete speech domain, aiming to generate discrete target speech tokens through a language modeling approach.
Compared to discriminative methods, generative methods generally produce speech with higher perceptual quality, as shown in \cite{difftse, generative, tang2024tselm}. However, these methods often overlook the intelligibility of the re-synthesized target speech, leading to semantic loss or alteration.

To address the aforementioned issues, we propose a joint training paradigm for generative TSE with both high perceptual quality and intelligibility.
Recent advances in target speaker ASR \cite{whisper_pt, whisper_sidecar, whisper_sq} demonstrate that the powerful encoder-decoder-based ASR model Whisper \cite{whisper}, despite being trained only on single-talker utterances, can be effectively prompted to focus on the target speaker in multi-talker overlapped utterances. Inspired by this capability, we adopt Whisper as the foundation model.
As illustrated in Fig. \ref{fig:main}, our model features a shared target speech encoder built on the pre-trained Whisper audio encoder, which is prompted to focus exclusively on the target speech using both a target speaker embedding and a segment of raw enrollment speech as cues. For efficient fine-tuning, we employ low-rank adaptation (LoRA) \cite{hu2021lora} within this module.
The target speech tokens extracted by the encoder are processed in a multi-task learning paradigm through two parallel branches: 1) an optimal-transport conditional flow matching (OT\text{-}CFM) module \cite{flow, mehta2024matcha, du2024cosyvoice} for synthesizing high-quality target speech; 2) a pre-trained text decoder to predict the target speech transcript as additional training supervision for enhanced intelligibility.
Experiments on well-established benchmarks Libri2Mix \cite{cosentino2020librimix} and WSJ0-2mix \cite{wsj} demonstrate that the proposed method extracts target speech in superior perceptual quality and intelligibility, highlighting the effectiveness of our proposed method.

\section{Method}
The proposed generative TSE model consists of three main modules: 1) a shared target speech encoder that extracts target speech tokens from the input speech mixture based on target speaker cues; 2) a flow-based token-to-spectrogram synthesizer that generates the target speech mel-spectrogram from the target speech tokens; 3) a text decoder that predicts the target speech transcript from the target speech tokens as additional training supervision to enhance speech intelligibility.

\subsection{Target Speech Encoder}

The target speech encoder is built on Whisper \cite{whisper}, a powerful ASR model trained on web-scale speech data. Specifically, the Whisper audio encoder takes a $d_f$-dimensional log-mel spectrogram $\mX \in \sR^{d_f \times T}$ as input and encodes it into $d_{m}$-dimensional hidden speech tokens $\mH \in \sR^{d_{m} \times T/2}$ as:
\begin{equation}
% \begin{split}
\label{eq1}
    {\mH} = {\rm AudioEncoder}_{\mathbb{\theta}_e}({\rm Conv}({\rm Pos}(\mX))),
    % \end{split}
\end{equation}
where the audio encoder is a multi-layer Transformer \cite{transformer} encoder parameterized by the pre-trained $\mathbb{\theta}_e$, and ${\rm Pos(\cdot)}$, ${\rm Conv(\cdot)}$ represent the positional encoding and convolution layer in the original Whisper implementation, respectively.

To adapt the pre-trained Whisper audio encoder into a target speech encoder, we propose a joint prompting scheme, prefixing both the target speaker embedding and the raw enrollment speech as in \cite{whisper_sidecar} ahead of the original model input.
% {Additionally, we apply LoRA-tuning \cite{hu2021lora} to the pre-trained model weights for efficient fine-tuning.}
LoRA-tuning \cite{hu2021lora} is further applied to the pre-trained model weights for efficient fine-tuning.
Specifically, given the target speaker embedding of the $i$-th speaker $\ve_i \in \sR^{d_e}$, the log-mel spectrogram of the enrollment speech $\mE_i \in \sR^{d_f \times T'}$, and the log-mel spectrogram of the multi-talker overlapped speech $\tilde{\mX} \in \sR^{d_f \times T}$, the target speech tokens $\mH_i \in \sR^{d_{m} \times T/2}$ are extracted as:
\begin{equation} \label{pAE}
\mH_i = {\rm AudioEncoder}_{\mathbb{\theta}_e'}({[\mW e_i, {\rm Conv}({[{\rm Pos'}(\mE_i), {\rm Pos}(\tilde{\mX})]}})]),
\end{equation}
where $[\cdot,\cdot]$ denotes concatenation along $T$, $\mW \in \sR^{d_m \times d_e}$ is an affine layer aligning the speaker embedding to the model dimension $d_m$, 
{and ${\rm Pos'(\cdot)}$ is an additional learnable positional encoding layer to distinguish the enrollment speech from the speech mixture input.}
% and ${\rm Pos'(\cdot)}$ is an additional positional encoding layer with extra learnable positional embeddings to distinguish the enrollment speech from the speech mixture input.
$\mathbb{\theta}_e'$ is LoRA-tuned model parameters as:
\begin{equation} \label{lora}
\mathbb{\theta}_e' = \{\mW_{0,l} + \mB_l\mA_l \mid \mW_{0,l} \in \mathbb{\theta}_{e,\text{LoRA}}\} \cup \mathbb{\theta}_{e,\text{others}},
\end{equation}
where \(\mathbb{\theta}_{e,\text{LoRA}}\) is the set of selected weights tuned via LoRA, specifically the weights in the \textit{query}, \textit{key}, \textit{value}, and \textit{output} layers of the \textit{attention} module. $\mB_l \in \sR^{d_m \times k}$ and $\mA_l \in \sR^{k \times d_m}$ are tunable low-rank matrices for the $l$-th selected weight matrix $\mW_{0,l}$.

% \subsection{Flow-Based Token-to-Spectrogram}
\subsection{{Flow-Based Token-to-Spectrogram Synthesizer}}
The mel-spectrogram of the target speech is re-synthesized from the target speech tokens by an optimal-transport conditional flow matching \cite{flow}, which is widely adopted in speech tasks \cite{mehta2024matcha, du2024cosyvoice, ning2024drop, yao2024stablevc} and has shown superior generation quality.
Formally, let \( \vx \in \mathbb{R}^d \) denotes an observation from an unknown distribution \( q(\vx) \). A probability density path is a time-dependent probability density function \( p_t : [0,1] \times \mathbb{R}^d \to \mathbb{R}>0 \). To generate samples from the data distribution \( q \), we can construct a probability density path \( p_t \), where \( t \in [0,1] \) and \( p_0(\vx) = \mathcal{N}(\vx; \bm{0}, \mI) \) is a prior distribution, such that \( p_1(\vx) \) approximates the data distribution \( q(\vx) \). Continuous normalizing flows define a vector field \( \vv_t : [0,1] \times \mathbb{R}^d \to \mathbb{R}^d \), which generates the flow \( \phi_t : [0,1] \times \mathbb{R}^d \to \mathbb{R}^d \) through the ordinary differential equation:

\begin{equation} \label{flow}
\frac{d}{dt} \phi_t(\vx) = \vv_t(\phi_t(\vx)); \quad \phi_0(\vx) = \vx.
\end{equation}

By solving Eq. \ref{flow}, we can approximate the speech distribution \( q(\vx) \) with \( p_1(\vx) \) and sample from it. To learn the vector field \( \vv_t(\vx) \), we adopt the optimal transport conditional flow \cite{flow} and force a neural network to match a conditional vector field by minimizing the following loss:
\begin{equation} \label{cfm}
\begin{aligned}
&\mathcal{L}_{OT\text{-}CFM} = \\
&\mathbb{E}_{t, p_0(\vx_0), q(\vx_1)} || \vu_t(\phi^{\text{OT}}_{t}(\vx) | \vx_1) - \vv_t(\phi^{\text{OT}}_{t}(\vx) | \mu;\theta_f) ||^2,
\end{aligned}
\end{equation}
where $\phi^{\text{OT}}_{t}(\vx) = (1 - (1 - \sigma)t) \vx_0 + t\vx_1$, $\vu_t(\phi^{\text{OT}}_{t}(\vx) | \vx_1) = \vx_1 - (1 - \sigma) \vx_0$, $\sigma$ is a hyperparameter with a small value, and $\mu$ represents a set of conditioning features, consisting of speech tokens $\mH$ and speaker embedding $\ve$. We follow the implementation in \cite{du2024cosyvoice} for the token-to-spectrogram module. Finally, the waveform of the target speech is re-synthesized from the mel-spectrogram by a HiFiGAN \cite{kong2020hifi} neural vocoder.

\subsection{Text Decoder}

The text decoder predicts the target speech transcript from the target speech tokens, providing additional training supervision to enhance speech intelligibility.
Formally, given the target speech tokens $\mH_i$ extracted by the target speech encoder, the text decoder predicts the probability of the next token conditioned on the previous decoding results as:
\begin{equation}
    p(y_{i,j}|[\vc, y_{i,1:j-1}], \mH_i)={\rm TextDecoder}_{\theta_d}([\vc, y_{i,1:j-1}], \mH_i),
\end{equation}
where $\vc$=[\textit{$\langle|$SOT$|\rangle$, $\langle|$EN$|\rangle$, $\langle|$transcribe$|\rangle$, $\langle|$no-timestamps$|\rangle$}] is a sequence of condition tokens used for decoding control, as defined in \cite{whisper}, and $\vy_i$ is a sequence of transcript tokens for the $i$-th target speech. The cross-entropy loss for the text decoder's prediction is calculated as:

\begin{equation} \label{ce}
\mathcal{L}_{CE} = - \sum_{j=1}^{N} \log \left( p(y_{i,j} | [\vc, y_{i,1:j-1}], \mH_i) \right).
\end{equation}

\subsection{Training Objectives}
The entire model is jointly optimized by minimizing a combination of the flow-matching loss $\mathcal{L}_{OT\text{-}CFM}$ and the cross-entropy loss $\mathcal{L}_{CE}$, resulting in the final loss function:
\begin{equation}
\mathcal{L} = \mathcal{L}_{OT\text{-}CFM}+\mathcal{L}_{CE}.
\end{equation}

For training efficiency, we only train the flow-based speech re-synthesizer, the LoRA matrices in the target speech encoder, and the positional embeddings for enrollment speech as detailed in Eq. \ref{pAE}. All other parameters are kept unchanged.

\section{Experiments}
\subsection{Datasets}
We use the training sets from the LibriSpeech \cite{librispeech} corpus, specifically \texttt{train\_clean\_100} and \texttt{train\_clean\_360}, to dynamically generate training speech mixtures. During training, two randomly selected utterances from different speakers are first zero-padded to match the same length. They are then mixed under a signal-to-noise ratio (SNR) sampled from a uniform distribution between (-5, 5) dB.
The enrollment utterance is chosen dynamically from the training set, ensuring it shares the same speaker label as the target speech, and it is padded or truncated to a fixed length of 5 seconds.
We use Libri2Mix \cite{cosentino2020librimix} and WSJ0-2mix \cite{wsj} as evaluation benchmarks. All evaluation speech mixtures are sampled at 16 kHz and constructed under the \textit{max} configuration as described in \cite{cosentino2020librimix}. For experiments on Libri2Mix, the enrollment speech is selected from a held-out dataset following prior works \cite{enroll, whisper_pt}. Each speech mixture for two speakers is extracted twice by alternating the enrollment utterances. For experiments on WSJ0-2mix, the enrollment speeches are selected as per \cite{xu2018single}.

\subsection{Evaluation Metrics}

{We employ a comprehensive set of metrics—deep noise suppression mean opinion score (DNSMOS) \cite{dnsmos}, word error rate (WER), SpeechBERTScore~\cite{saeki2024speechbertscore}, and cosine similarity—to evaluate the perceptual quality, intelligibility, and timbre consistency of the extracted target speech, respectively.} DNSMOS is primarily designed to assess speech perceptual quality based on a deep learning model, but lacks a focus on speech intelligibility. 
{We therefore report two metrics as a comprehensive evaluation for speech intelligibility: the WER, computed by comparing the transcription produced by a pre-trained ASR model (Whisper-small) on the TSE model’s output speech with the ground-truth text; and the SBS, derived by comparing semantic embeddings extracted from the TSE model’s output speech with those extracted from the reference speech.}
% which is calculated by comparing transcripts generated by a pre-trained ASR model (Whisper-small) against the ground truth.
Additionally, cosine similarity is used as a general indicator of non-lingual information preservation by comparing the embeddings of the model’s predictions with those of the reference speech. The embeddings are extracted using a pre-trained CAM++ \cite{cam} model.
Note that, due to the compression of high-frequency components and the absence of phase in generative methods involving mel-based vocoders, intrusive metrics such as SNR are not applicable for evaluating the performance of such methods \cite{generative, tang2024tselm, selm}.

\subsection{Implementation Details}

We implement three different-sized models based on Whisper Small, Medium, and Large-V3, referred to as WhisperTSE-S, -M, and -L, respectively. For efficiency, we initialize the flow-based token-to-spectrogram module from pre-trained weights provided by CosyVoice \cite{du2024cosyvoice} instead of training it from scratch. All models are trained on GPU accelerator(s) using distributed-data-parallel with a global batch size of 16. The training is conducted with the AdamW \cite{adamw} optimizer over a total of 10 epochs, starting with an initial learning rate of 1e-4, which is reduced by a factor of 0.1 at epoch 5.

\subsection{Results and Analysis}

\begin{figure}
    \centering
    \includegraphics[width=0.8\linewidth]{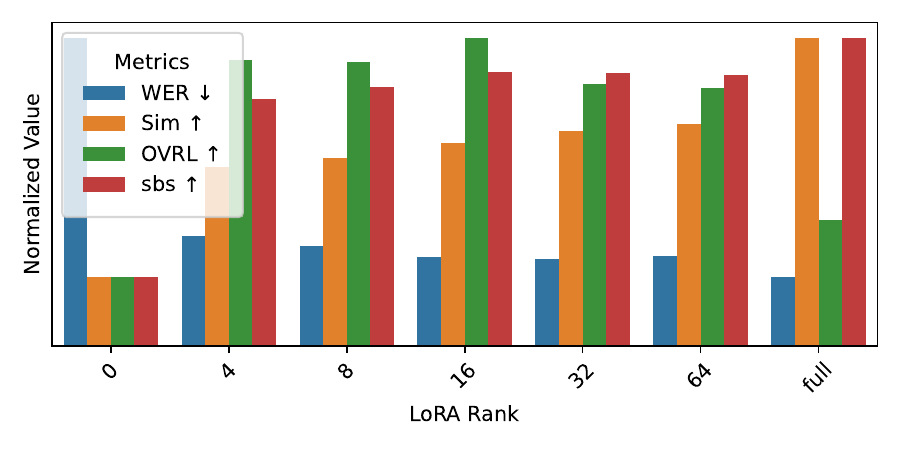}
    \vspace{-3.5mm}
    \caption{Normalized metrics versus LoRA rank. All metrics are evaluated with WhisperTSE-S on \texttt{Libri2Mix-dev-clean}.}
    \label{fig:rank}
    \vspace{-5mm}
\end{figure}

\begin{table*}[t]
        \scriptsize
	\caption{Results on target speech extraction. $^\dag$: implemented with CAM++ \cite{cam} as speaker embedder and BSRoformer \cite{roformer} as target speaker extractor following \cite{pbsrnn}. $^\ddag$: implemented by cascading the text decoder output with a text-to-speech backend \cite{du2024cosyvoice}.} %标题
	\centering
	\label{tab:tse}
	\begin{tabular}{lccccccccccccc}
		\toprule
		\multirow{3}*{Method}  & \multirow{3}*{Category} & \multicolumn{6}{c}{Libri2Mix} & \multicolumn{6}{c}{WSJ0-2mix}\\
		\cmidrule(lr){3-8} \cmidrule(lr){9-14}
            &&\multicolumn{3}{c}{DNSMOS$\uparrow$}&\multirow{2}*{WER$\downarrow$}&\multirow{2}*{{SBS$\uparrow$}}&\multirow{2}*{Cos Sim$\uparrow$}&\multicolumn{3}{c}{DNSMOS$\uparrow$}&\multirow{2}*{WER$\downarrow$}&\multirow{2}*{{SBS$\uparrow$}}&\multirow{2}*{Cos Sim$\uparrow$}\\
		  \cmidrule(lr){3-5} \cmidrule(lr){9-11}
          &&SIG&BAK&OVL&&&&SIG&BAK&OVL\\
          \midrule
		
		Mixture  & - & 3.521 & 3.530 & 2.978 & 0.703&{0.702}& 0.668&3.535&3.493&3.002&0.457&{0.732}&0.739  \\
		Clean& - & 3.558 & 4.021 & 3.248 & 0.034 &{-}& - &3.607&4.009&3.309&0.044&{-}&-\\
        \midrule
        \midrule
        % X-TF-GridNet&D& \\
        pBSRoformer$^\dag$&D&3.475&3.904&3.131&0.106&{0.879}&\textbf{0.905}&3.506&3.890&3.169&0.083&{0.886}&\textbf{0.927}\\
        TSELM-L \cite{tang2024tselm} &G&3.553&4.121&3.299&0.312&{0.801}&0.563&3.597&4.111&3.343&0.152&{0.823}&0.581 \\
        \midrule
        \textcolor{gray}{Cascading$^\ddag$}&\textcolor{gray}{G}&\textcolor{gray}{3.632}&\textcolor{gray}{4.096}&\textcolor{gray}{3.362}&\textcolor{gray}{0.070}&\textcolor{gray}{0.829}&\textcolor{gray}{0.702}&\textcolor{gray}{3.610}&\textcolor{gray}{4.027}&\textcolor{gray}{3.320}&\textcolor{gray}{0.080}&\textcolor{gray}{0.840}&\textcolor{gray}{0.810}\\
        \midrule
        WhisperTSE-S&G&3.614&4.162&3.377&0.142&{0.907}&0.749&3.617&4.160&3.392&0.084&{0.915}&0.814\\
        WhisperTSE-M&G& {\bf3.619}&\textbf{4.164}&\textbf{3.384}&0.109&{0.914}&0.762&\textbf{3.624}&4.166&\textbf{3.402}&0.074&{0.917}&0.824\\
        WhisperTSE-L&G& 3.616&4.163&3.379&{\bf 0.081}&{\bf 0.927}& 0.783&{\bf 3.624}&\textbf{4.167}&\textbf{3.402}&\textbf{0.062}&{\bf 0.927}&0.846\\
        
        % avg.   & 14.26 & 13.44 & {\bf 13.03} & 13.50 & 13.17& - \\
		\bottomrule
	\end{tabular}
    % }
 \vspace{-5mm}
\end{table*}

\begin{table}[t]
	\caption{Ablation study. All experiments are conducted on \texttt{Libri2Mix-test-clean} using the WhisperTSE-S model. %\textit{``w/o enroll emb''} denotes we only provide the raw enrollment speech for the target speech encoder as target speaker cues, and \textit{``w/o enroll speech''} denotes we only provide the target speaker embedding for the target speech encoder.
    } %标题
	\centering
        \scriptsize
	\label{tab:ablation}
    \setlength{\tabcolsep}{0.5mm}{
	\begin{tabular}{lcccccc}
		\toprule
		\multirow{2}*{Method}
            &\multicolumn{3}{c}{DNSMOS$\uparrow$}&\multirow{2}*{WER$\downarrow$}&\multirow{2}*{SBS$\uparrow$}&\multirow{2}*{Cos Sim$\uparrow$}\\
		  \cmidrule(lr){2-4} 
          &SIG&BAK&OVL&&\\
          \midrule
		
		Ours best & 3.614& 4.162 & 3.377 & 0.142&0.907& 0.749  \\
		\rowcolor{gray!30}{\it \quad - w/o spk emb}&3.608&4.153& 3.369 & 0.143&0.906  & 0.749 \\
            {\it \quad - w/o enroll speech}&3.606&4.152& 3.366 & 0.220&  0.883& 0.722 \\
        \rowcolor{gray!30}{\it \quad - w/o joint training}&3.606&4.144& 3.360 & 0.234&0.891& 0.793 \\
            % {\it \quad - w/o pre-trained flow}&3.474&4.054& 3.195 & 0.148& 0.879 & 0.695 \\
        % \rowcolor{gray!30}{\it \qquad -  replace flow with diffusion}&3.405&4.003& 3.124 & 0.156&0.866 & 0.702 \\
        
        % avg.   & 14.26 & 13.44 & {\bf 13.03} & 13.50 & 13.17& - \\
		\bottomrule
	\end{tabular}}
 \vspace{-5mm}
\end{table}

% \begin{table}[t]
% 	\caption{Results on target speaker ASR. All experiments are conducted on \texttt{Libri2Mix-test-clean}. $\dag$: trained on a limited dataset \texttt{Libri2Mix-train-100} w/o dynamic mixing. S, M, and L denote Whisper small, medium, and large respectively; $\ddag$: large-v2 in \cite{} and -v3 in \cite{} and ours.} %标题
% 	\centering
% 	\label{tab:tsasr}
%     \setlength{\tabcolsep}{1.6mm}{
% 	\begin{tabular}{lccc}
% 		\toprule
% 		\multirow{2}*{Method}  & \multicolumn{3}{c}{WER (\%) $\downarrow$}\\
% 		\cmidrule(lr){2-4}
%           &S&M&L$^\ddag$\\
%           \midrule
%           Mixture &70.28&64.79&65.20\\
%           Clean&3.41&2.87&1.92\\

%           \midrule
		
% 		Whisper-LoRA$^\dag$ \cite{}  & 19.06 & 11.98 & 13.07  \\
% 		Whisper-Sidecar \cite{} &11.81&9.14&7.97 \\
%             Ours &9.46&6.19&4.90\\
        
%         % avg.   & 14.26 & 13.44 & {\bf 13.03} & 13.50 & 13.17& - \\
% 		\bottomrule
% 	\end{tabular}}
%  % \vspace{-7mm}
% \end{table}

\subsubsection{Selection of LoRA Rank}
{
We first evaluate the LoRA rank setting by conducting experiments with our proposed \textit{WhisperTSE-S} model on \texttt{Libri2Mix-dev-clean}. Specifically, we set the LoRA rank to \textit{\{0, 4, 8, 16, 32, 64, full\}}, where \textit{0} means no LoRA fine-tuning and \textit{full} means full fine-tuning. For each rank setting, we trained a WhisperTSE-S model and then evaluated each model on the \texttt{Libri2Mix-dev-clean}; all results are shown in Fig. \ref{fig:rank}. As the figure shows, LoRA fine-tuning yields a significant performance boost, and a rank of 16 offers the best trade-off between model performance and training cost. Therefore, in all subsequent experiments, we fix the LoRA rank at 16.
}
\subsubsection{Results on Target Speech Extraction}

% We first show the proposed model's performance on target speech extraction and compare it with other generative and discriminative baselines. 
{We demonstrate the performance of the proposed model in target speech extraction and compare it with other generative and discriminative baselines.}
% For comparison, we introduce a mask-based discriminative baseline called pBSRoformer, which is implemented with CAM++ \cite{cam} as a speaker embedding module and BSRoformer \cite{roformer}, a state-of-the-art source separation model, as a target speech extractor. 
{For comparison, we introduce a mask-based discriminative baseline called \textit{pBSRoformer}, which uses CAM++ \cite{cam} as the speaker embedding module and the state-of-the-art source separation model BSRoformer \cite{roformer} as the target speech extractor.}
This model is trained on the same dataset as our proposed WhisperTSE. Additionally, we introduce two generative baselines: one is \textit{TSELM} \cite{tang2024tselm}, which models target speech in the discrete domain, and the other is a vanilla cascading pipeline, where the output of the text decoder is directly cascaded with a text-to-speech backend CosyVoice \cite{du2024cosyvoice}. All results are presented in Table \ref{tab:tse}.

Compared with the discriminative baseline pBSRoformer, generative methods avoid the over-/under-suppression issues caused by imperfect mask modeling, resulting in better perceptual quality, as indicated by the DNSMOS metrics. Moreover, the discriminative baseline does not explicitly model the semantic information in the speech, leading to poorer intelligibility compared to our method. 
%This highlights the superiority of the proposed approach, which leverages advanced flow-based generative techniques and explicit semantic modeling.
The proposed generative method falls short of the discriminative method in terms of cosine similarity. This is primarily due to the generative model's output not being strictly aligned with the target at the sample level, a common observation also reported in other tasks such as generative speech enhancement \cite{selm}.

Compared with the discrete-domain generative baseline TSELM \cite{tang2024tselm}, our proposed method comprehensively outperforms it in terms of perceptual quality, intelligibility, and timbre consistency. This demonstrates the effectiveness of our approach in modeling continuous speech tokens through optimal-transport conditional flow matching for improved perceptual quality, as well as the advantage of leveraging Whisper for joint acoustic and semantic modeling to enhance speech intelligibility.
Finally, we compare against a vanilla cascading baseline. Although it performs slightly better than the proposed method in terms of intelligibility on Libri2Mix, the cascading approach—where the speech mixture is transcribed into target text and then used to synthesize target speech—loses all the rich non-linguistic information present in the original speech mixture. In contrast, our method retains much of this rich paralinguistic information, as evidenced by the superiority indicated by the cosine similarity metric.

\subsubsection{Ablation Experiments}

We then conduct an ablation study to evaluate the effectiveness of each design component. All results are shown in Table \ref{tab:ablation}.
First, we explore the effectiveness of various prompting strategies in adapting the pre-trained Whisper audio encoder into a target speech encoder by ablating the target speaker embedding (\textit{``w/o spk emb''}) or raw enrollment speech (\textit{``w/o enroll speech''}). Overall, the proposed joint prompting scheme achieves optimal performance by effectively integrating both the target speaker embedding and the raw enrollment speech.
It is worth noting that raw enrollment speech plays a crucial role, as evidenced by a significant performance deterioration after its ablation. This highlights the importance of the contextual information contained in the raw enrollment speech for accurately identifying the target speaker by prompting the whisper model, a finding also reported in recent studies \cite{zhang2024multi, multi_1}.
Importantly, the proposed joint optimization approach leads to a significant improvement in speech intelligibility, as evidenced by the significant deterioration of WER and SBS metrics after its ablation. The improvement in cosine similarity after the ablation of joint training suggests that the model is overly focused on acoustic details, leading to a loss in its semantic modeling capability. %{We further validate the importance of the proposed flow-based token-to-spectrogram module by replacing it with a diffusion-based generative module, following the implementation in~\cite{grad}. Notably, the pre-trained weights of the flow-based module are crucial for achieving high-quality speech generation. To ensure a fair comparison, we first ablate the pre-trained weights of the flow-based module (\textit{``w/o pre-trained flow''}). Then we replace the entire flow-based module with a diffusion-based module (\textit{``replace flow with diffusion''}). As shown in the last two rows of Table~\ref{tab:ablation}, the flow-based module consistently outperforms the diffusion-based counterpart.}

% \subsubsection{Results on Target Speaker ASR}

% We evaluate the performance of the proposed method on the target speaker ASR task, which is a byproduct of the joint training strategy. As shown in Table \ref{tab:tsasr}, the proposed method also outperforms existing state-of-the-art target speaker ASR models in this task.
% This demonstrates the effectiveness of the proposed method in modeling the target speaker's speech within the speech mixture.

\subsubsection{Visualization Analysis}

\begin{figure}
    \centering
    \includegraphics[width=0.7\linewidth]{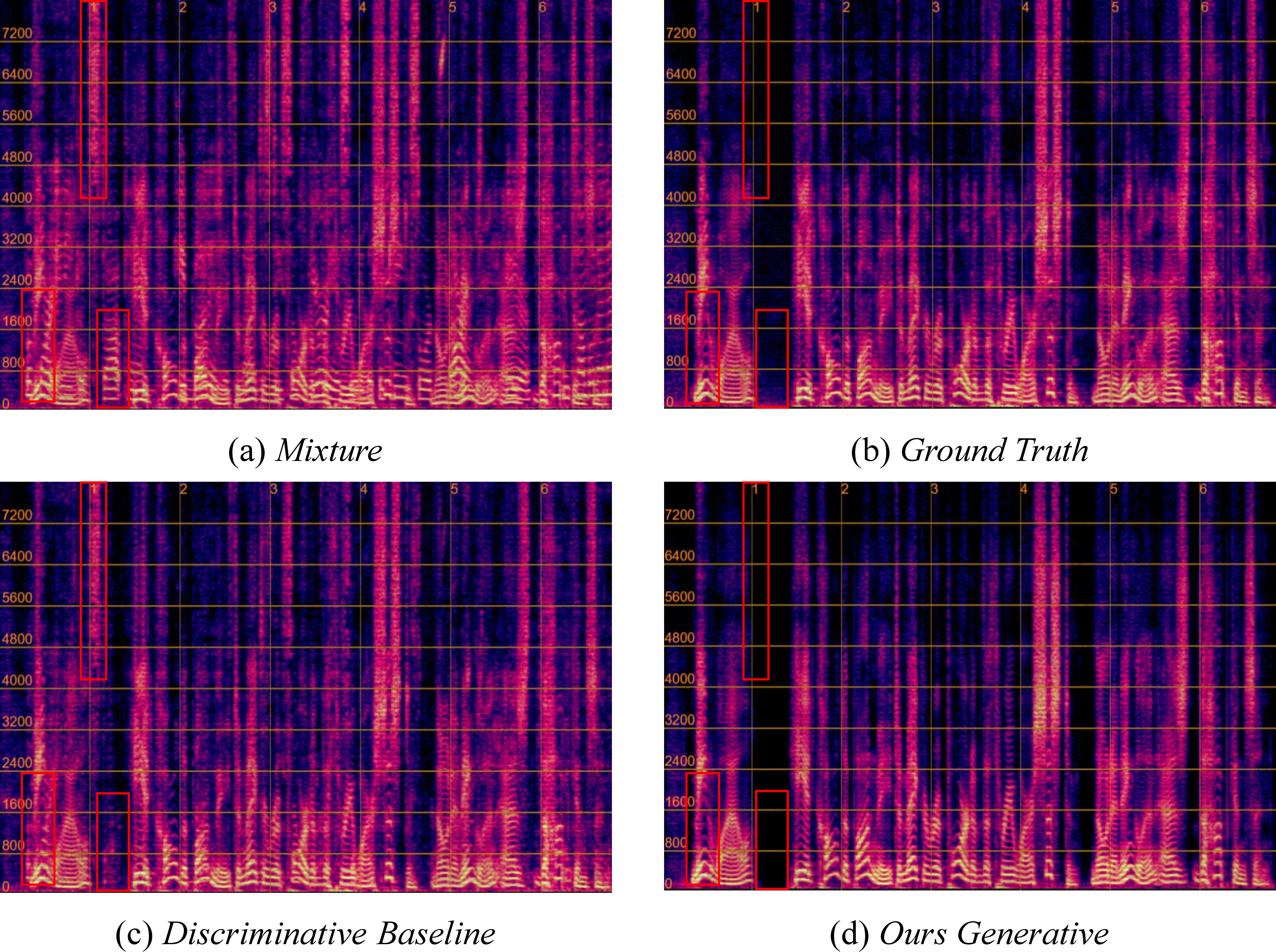}
    \vspace{-2mm}
    \caption{Spectrogram Visualization}
    \label{fig:vis}
    \vspace{-5mm}
\end{figure}

We visualize the spectrograms of predictions made by our proposed method and the discriminative baseline in Fig. \ref{fig:vis}.
The spectrograms produced by the proposed generative method are noticeably cleaner, demonstrating superior perceptual quality. Moreover, the speech extracted using our model, based on its energy distribution in the time-frequency spectrogram, shows no perceptible distortion compared to the ground truth. This suggests that both the linguistic and non-linguistic information in the original speech mixture is effectively preserved.

\section{Discussion}

In this work, we propose a generative target speech extraction framework based on the pre-trained Whisper model, integrating semantic modeling with optimal-transport conditional flow matching-based acoustic modeling to achieve both high perceptual quality and intelligibility. Multiple evaluations on well-established benchmarks demonstrate the effectiveness of our proposed method. 
{In particular, our framework draws attention to previously overlooked dimensions of generative TSE and calls for a comprehensive evaluation paradigm that jointly measures perceptual fidelity and speech intelligibility.}
However, although experiments show that the speech extracted by the proposed method outperforms traditional discriminative methods in terms of speech quality, the decrease in cosine similarity suggests that our generative approach still introduces some distortions, leading to changes or loss of non-linguistic details. This issue has also been reported when applying generative methods to other tasks, such as speech enhancement. Finding solutions to this problem remains an area for further exploration. {Also, the introduction of a pre-trained Whisper model into the TSE framework causes additional inference cost, which may be prohibitive for real-time applications. While preserving the speech intelligibility benefits provided by the pre-trained Whisper model, applying lightweighting techniques such as pruning to reduce additional inference overhead is a promising direction for further research.}

% \section*{Acknowledgment}

% The preferred spelling of the word ``acknowledgment'' in American English is without an ``e'' after the ``g.'' Use the singular heading even if you have many acknowledgments. Avoid expressions such as ``One of us (S.B.A.) would like to thank . . . .'' Instead, write “F. A. Author thanks ... .” In most cases, sponsor and financial support acknowledgments are placed in the unnumbered footnote on the first page, not here.

\bibliographystyle{IEEEtran}
\bibliography{IEEEabrv,mybib}

\begin{thebibliography}{10}
\providecommand{\url}[1]{#1}
\csname url@samestyle\endcsname
\providecommand{\newblock}{\relax}
\providecommand{\bibinfo}[2]{#2}
\providecommand{\BIBentrySTDinterwordspacing}{\spaceskip=0pt\relax}
\providecommand{\BIBentryALTinterwordstretchfactor}{4}
\providecommand{\BIBentryALTinterwordspacing}{\spaceskip=\fontdimen2\font plus
\BIBentryALTinterwordstretchfactor\fontdimen3\font minus \fontdimen4\font\relax}
\providecommand{\BIBforeignlanguage}[2]{{%
\expandafter\ifx\csname l@#1\endcsname\relax
\typeout{** WARNING: IEEEtran.bst: No hyphenation pattern has been}%
\typeout{** loaded for the language `#1'. Using the pattern for}%
\typeout{** the default language instead.}%
\else
\language=\csname l@#1\endcsname
\fi
#2}}
\providecommand{\BIBdecl}{\relax}
\BIBdecl

\bibitem{SS}
D.~Wang and J.~Chen, ``Supervised speech separation based on deep learning: An overview,'' \emph{IEEE/ACM Trans. Audio, Speech, Lang. Process.}, vol.~26, no.~10, pp. 1702--1726, 2018.

\bibitem{speakerbeam_lstm}
K.~Žmolíková, M.~Delcroix, K.~Kinoshita, T.~Ochiai, T.~Nakatani, L.~Burget, and J.~Černocký, ``{SpeakerBeam}: Speaker aware neural network for target speaker extraction in speech mixtures,'' \emph{IEEE J. Sel. Topics Signal Process.}, vol.~13, no.~4, pp. 800--814, 2019.

\bibitem{speakerbeam_cnn}
M.~Delcroix, T.~Ochiai, K.~Zmolikova, K.~Kinoshita, N.~Tawara, T.~Nakatani, and S.~Araki, ``Improving speaker discrimination of target speech extraction with time-domain speakerbeam,'' in \emph{Proc. ICASSP}, 2020, pp. 691--695.

\bibitem{xu2020spex}
C.~Xu, W.~Rao, E.~S. Chng, and H.~Li, ``{SpEx}: Multi-scale time domain speaker extraction network,'' \emph{IEEE/ACM Trans. Audio, Speech, Lang. Process.}, vol.~28, pp. 1370--1384, 2020.

\bibitem{difftse}
N.~Kamo, M.~Delcroix, and T.~Nakatani, ``Target speech extraction with conditional diffusion model,'' in \emph{Proc. INTERSPEECH}, 2023, pp. 176--180.

\bibitem{generative}
L.~Yu, W.~Zhang, C.~Du, L.~Zhang, Z.~Liang, and Y.~Qian, ``Generation-based target speech extraction with speech discretization and vocoder,'' in \emph{Proc. ICASSP}, 2024, pp. 12\,612--12\,616.

\bibitem{tang2024tselm}
B.~Tang, B.~Zeng, and M.~Li, ``{TSELM}: Target speaker extraction using discrete tokens and language models,'' \emph{arXiv preprint arXiv:2409.07841}, 2024.

\bibitem{xsep}
K.~Liu, Z.~Du, X.~Wan, and H.~Zhou, ``{X-Sepformer}: End-to-end speaker extraction network with explicit optimization on speaker confusion,'' in \emph{Proc. ICASSP}, 2023, pp. 1--5.

\bibitem{pbsrnn}
X.~Le, L.~Chen, C.~He, Y.~Guo, C.~Chen, X.~Xia, and J.~Lu, ``Personalized speech enhancement combining band-split rnn and speaker attentive module,'' in \emph{Proc. ICASSP}, 2023, pp. 1--2.

\bibitem{zhang2024multi}
K.~Zhang, J.~Li, S.~Wang, Y.~Wei, Y.~Wang, Y.~Wang, and H.~Li, ``Multi-level speaker representation for target speaker extraction,'' \emph{arXiv preprint arXiv:2410.16059}, 2024.

\bibitem{multi_1}
S.~He, H.~Zhang, W.~Rao, K.~Zhang, Y.~Ju, Y.~Yang, and X.~Zhang, ``Hierarchical speaker representation for target speaker extraction,'' in \emph{Proc. ICASSP}, 2024, pp. 10\,361--10\,365.

\bibitem{whisper_pt}
H.~Ma, Z.~Peng, M.~Shao, J.~Li, and J.~Liu, ``{Extending {Whisper} with prompt tuning to target-speaker ASR},'' in \emph{Proc. ICASSP}, 2024.

\bibitem{whisper_sidecar}
L.~Meng, J.~Kang, Y.~Wang, Z.~Jin, X.~Wu, X.~Liu, and H.~Meng, ``{Empowering {Whisper} as a joint multi-talker and target-talker speech recognition system},'' in \emph{Proc. INTERSPEECH}, 2024.

\bibitem{whisper_sq}
P.~Guo, X.~Chang, H.~Lv, S.~Watanabe, and L.~Xie, ``{SQ-{Whisper}}: Speaker-querying based {Whisper} model for target-speaker {ASR},'' \emph{IEEE/ACM Trans. Audio, Speech, Lang. Process.}, pp. 1--11, 2024.

\bibitem{whisper}
A.~Radford, J.~W. Kim, T.~Xu, G.~Brockman, C.~McLeavey, and I.~Sutskever, ``Robust speech recognition via large-scale weak supervision,'' in \emph{Proc. ICML}, 2023, pp. 28\,492--28\,518.

\bibitem{hu2021lora}
E.~J. Hu, Y.~Shen, P.~Wallis, Z.~Allen-Zhu, Y.~Li, S.~Wang, L.~Wang, and W.~Chen, ``Lo{RA}: Low-rank adaptation of large language models,'' \emph{ICLR}, 2022.

\bibitem{flow}
Y.~Lipman, R.~T. Chen, H.~Ben-Hamu, M.~Nickel, and M.~Le, ``Flow matching for generative modeling,'' \emph{ICLR}, 2023.

\bibitem{mehta2024matcha}
S.~Mehta, R.~Tu, J.~Beskow, {\'E}.~Sz{\'e}kely, and G.~E. Henter, ``Matcha-{TTS}: A fast {TTS} architecture with conditional flow matching,'' in \emph{Proc. ICASSP}, 2024.

\bibitem{du2024cosyvoice}
Z.~Du, Q.~Chen, S.~Zhang, K.~Hu, H.~Lu, Y.~Yang, H.~Hu, S.~Zheng, Y.~Gu, Z.~Ma \emph{et~al.}, ``{CosyVoice}: A scalable multilingual zero-shot text-to-speech synthesizer based on supervised semantic tokens,'' \emph{arXiv preprint arXiv:2407.05407}, 2024.

\bibitem{cosentino2020librimix}
J.~Cosentino, M.~Pariente, S.~Cornell, A.~Deleforge, and E.~Vincent, ``{LibriMix}: An open-source dataset for generalizable speech separation,'' 2020.

\bibitem{wsj}
J.~R. Hershey, Z.~Chen, J.~Le~Roux, and S.~Watanabe, ``Deep clustering: Discriminative embeddings for segmentation and separation,'' in \emph{Proc. ICASSP}, 2016, pp. 31--35.

\bibitem{transformer}
A.~Vaswani, N.~Shazeer, N.~Parmar, J.~Uszkoreit, L.~Jones, A.~N. Gomez, {\L}.~Kaiser, and I.~Polosukhin, ``Attention is all you need,'' in \emph{Proc. NeurIPS}, vol.~30, 2017.

\bibitem{ning2024drop}
Z.~Ning, S.~Wang, Y.~Jiang, J.~Yao, L.~He, S.~Pan, J.~Ding, and L.~Xie, ``Drop the beat! freestyler for accompaniment conditioned rapping voice generation,'' \emph{arXiv preprint arXiv:2408.15474}, 2024.

\bibitem{yao2024stablevc}
J.~Yao, Y.~Yan, Y.~Pan, Z.~Ning, J.~Ye, H.~Zhou, and L.~Xie, ``{StableVC}: Style controllable zero-shot voice conversion with conditional flow matching,'' \emph{arXiv preprint arXiv:2412.04724}, 2024.

\bibitem{kong2020hifi}
J.~Kong, J.~Kim, and J.~Bae, ``{HiFi-GAN}: Generative adversarial networks for efficient and high fidelity speech synthesis,'' in \emph{Proc. NeurIPS}, vol.~33, 2020, pp. 17\,022--17\,033.

\bibitem{librispeech}
V.~Panayotov, G.~Chen, D.~Povey, and S.~Khudanpur, ``{LibriSpeech}: An {ASR} corpus based on public domain audio books,'' in \emph{Proc. ICASSP}, 2015, pp. 5206--5210.

\bibitem{enroll}
Z.~Huang, D.~Raj, P.~García, and S.~Khudanpur, ``Adapting self-supervised models to multi-talker speech recognition using speaker embeddings,'' in \emph{Proc. ICASSP}, 2023.

\bibitem{xu2018single}
C.~Xu, W.~Rao, X.~Xiao, E.~S. Chng, and H.~Li, ``Single channel speech separation with constrained utterance level permutation invariant training using grid lstm,'' in \emph{Proc. ICASSP}, 2018, pp. 6--10.

\bibitem{dnsmos}
C.~K.~A. Reddy, V.~Gopal, and R.~Cutler, ``{DNSMOS P.835}: A non-intrusive perceptual objective speech quality metric to evaluate noise suppressors,'' in \emph{Proc. ICASSP}, 2022, pp. 886--890.

\bibitem{saeki2024speechbertscore}
T.~Saeki, S.~Maiti, S.~Takamichi, S.~Watanabe, and H.~Saruwatari, ``Speechbertscore: Reference-aware automatic evaluation of speech generation leveraging nlp evaluation metrics,'' \emph{arXiv preprint arXiv:2401.16812}, 2024.

\bibitem{cam}
H.~Wang, S.~Zheng, Y.~Chen, L.~Cheng, and Q.~Chen, ``{CAM++}: A fast and efficient network for speaker verification using context-aware masking,'' in \emph{Proc. INTERSPEECH}, 2023, pp. 5301--5305.

\bibitem{selm}
Z.~Wang, X.~Zhu, Z.~Zhang, Y.~Lv, N.~Jiang, G.~Zhao, and L.~Xie, ``{SELM}: Speech enhancement using discrete tokens and language models,'' in \emph{Proc. ICASSP}, 2024, pp. 11\,561--11\,565.

\bibitem{adamw}
I.~Loshchilov and F.~Hutter, ``Decoupled weight decay regularization,'' \emph{ICLR}, 2019.

\bibitem{roformer}
W.-T. Lu, J.-C. Wang, Q.~Kong, and Y.-N. Hung, ``Music source separation with band-split rope transformer,'' in \emph{Proc. ICASSP}, 2024, pp. 481--485.

\end{thebibliography}

\end{document}